\documentclass{article}
\usepackage{epsfig}
\usepackage{amsmath,amssymb}

\usepackage{graphicx}

\def\be{\begin{equation}}
\def\ee{\end{equation}}
\def\bea{\begin{eqnarray}}
\def\eea{\end{eqnarray}}
\def\ba{$$\begin{array}}
 \def\ea{\end{array}$$}

\newcommand{\dst}{\displaystyle}

\newcommand{\fr}[2]{\frac{{\dst #1}}{{\dst #2}}}

\def\Pom{{\bf I\!P}}
\def\od{\mathbb{O}}
\def\cl{\centerline}
\def\bu{$\bullet$ }
\newcommand{\bear}{\be\begin{array}}
{\end{list}}
\newcounter{enumct}

\newsavebox{\fmbox}
    
\date{}

\begin{document}

\title{\Large\bf How to measure Pomeron phase and  discover odderon at  HERA
and RHIC }
\author{I. F. Ginzburg$^1$, I.P. Ivanov$^{1,2}$\\
$^1$ Sobolev Inst. of Mathematics,  Novosibirsk, Russia\\
$^2$ INFN, Gruppo collegato di Cosenza, Italy}

 \maketitle

{\large\it To be published in Proceedings "Photon2005",
Acta
Physica Polonica}\\

\begin{abstract}
We suggest to measure Pomeron phase and discover odderon
via the measurement of charge asymmetry of pions in the
diffractive processes $ep\to e\pi^+\pi^- p$, $eA\to
e\pi^+\pi^- A$ and in the processes $AA\to AA\pi^+\pi^-$
with two rapidity gaps.
\end{abstract}

To measure the Pomeron ($\Pom$) phase in $\gamma p$, $\gamma A$
reactions and to discover the odderon ($\od$) we suggest to
study {\it charge asymmetry of pions} with small transverse
momentum $k_\bot = (p_++p_-)_\bot$  of a dipion with
effective mass $M=1.1\div 1.4$~GeV in the processes with
two rapidity gaps $ep\to e\pi^+\pi^-p$ at HERA ($\Pom$ and
$\od$), $eA\to e\pi^+\pi^-A$ at e-RHIC ($\Pom$ and $\od$),
$AA\to A\pi^+\pi^-A$  at RHIC (only $\Pom$). This charge
asymmetry is due to the interference of the C--odd and C--even
dipion production amplitudes. Small
values of $k_\bot$ are obtained at reasonable values of
measurable pion transverse momenta $p_{\pm \bot}\sim
400\div 500$~MeV providing good observability of the effects.
This charge asymmetry can be seen even if the process with
production of C--even dipion is below observation limit.\\

$\blacksquare$ {\Large\bf Motivation}

The  Pomeron and  odderon are both the $t$--channel objects
for hadronic $2\to 2$ processes with vacuum quantum numbers
and the only difference:  the Pomeron is $C$--even, while
the odderon is $C$--odd (similarly to the photon).

At high energies the Pomeron  amplitude $A_\Pom\propto
s^{\alpha_\Pom}$ describes small angle elastic scattering
and total cross sections. The odderon  amplitude
$A_\od\propto s^{\alpha_\od}$ describes difference of
hadronic ($h$) cross sections $\sigma_{hh}-\sigma_{h\bar
h}$ \cite{Luk}  and small angle cross sections of processes
\cite{Zhit} $d\sigma(\gamma p\to f_2p)$, $d\sigma(\gamma
p\to\pi^0 p)$,... . Here $\alpha_\Pom$, $\alpha_\od$ are
the Pomeron and odderon intercepts. Within perturbative
QCD, the Pomeron and odderon are based on two--gluon and
$d$--coupled three--gluon exchanges in $t$--channel
\cite{Bart} meaning $\alpha_\od,\, \alpha_\Pom \sim 1$
(gluon spin). (For more details and history see
\cite{Gin2001,Ginchas,Gin2004}.)

\bu The phase $\delta_F$ of the forward   high energy
hadronic elastic scattering amplitude $A$ (Pomeron phase)
is given  by
 $ {\cal A}=|{\cal A}|e^{i\delta_F}\equiv |{\cal A}|
\exp\left[i \pi(1+\Delta_F)/2\right]
 $.

Each model has its own characteristic process-dependence of
$\Delta_F$.\\ $\blacktriangledown$ In  the naive Regge-pole
Pomeron model
$\Delta_F=-(\alpha_\Pom-1)$,\\
$\blacktriangledown$ in the model of a dipole Pomeron
$\Delta_F=-(\alpha_\Pom -1) - \pi/(2\ln (s/s_0))$,\\
$\blacktriangledown$  for the model with Regge pole and
cuts $\Delta_F=-(\alpha_\Pom-1) +$ process dependent
contribution of the branch cuts.

The only reaction where such phase has been measured --
high energy elastic $pp$, $\bar{p}p$ scattering near
forward direction -- involves detailed measurement of the
cross section at extremely low transverse momentum of
recorded particle, $p_\bot\approx\sqrt{|t|} \lesssim 50$
MeV (extremely small scattering angles) (see e.g.
\cite{pp}). This will be a very difficult task at LHC.

\bu The  odderon is necessary but elusive  element of the
QCD motivated hadron physics. The odderon-induced
asymptotic difference $\sigma_{pp}-\sigma_{p\bar p}$ can be
zero within experimental uncertainty. The attempts to
discover the odderon via $\gamma p\to\pi^0 p'$, $\gamma
p\to f_2 p'$ at HERA  \cite{oddexpHERA} were based on
calculation of ref.~\cite{Berger} (containing inaccuracies,
see \cite{Gin2001} for details). New estimates of the same
group  \cite{Dosh1} lie below the upper experimental
limits. A reanalysis of HERA data on $\gamma p\to f_2p'$,
etc. is necessary due to inaccuracies of previous
theoretical estimates.

In our analysis we estimate the observability the odderon
signal in $\gamma p\to f_2p$ if its cross section is larger
than 1~nb (0.03 from upper limit given
by experiment \cite{oddexpHERA}).\\

$\blacksquare$ {\Large\bf Notation}

We consider kinematics of process having in mind $\gamma p\to \pi^+\pi^-p$
or $\gamma A\to \pi^+\pi^-A$ subprocess.
We denote dipion momentum as $k=p_-+p_+$, where $p_\pm$ is
momentum of $\pi^\pm$,
$M\equiv \sqrt{k^2}$, $\beta = \sqrt{1-4m_\pi^2/M^2}$,
$z_\pm= (\epsilon_\pm+p_{\pm z})/(2E_\gamma)=(p_\pm P)/(qP)$, $J$
and $\lambda_{\pi\pi}$ are spin and helicity of dipion,
$\lambda_\gamma$ is the photon helicity. Besides we define
some angles in the dipion c.m.s by relation
$(p_- - p_+)_{cms}= \beta M(0,\,\sin\theta\cos\phi\,,\;\sin\theta\cos\phi\,,\; cos\theta)$.
We describe  {\bf forward--backward (FB) and
transverse (T) asymmetries} by variables\
 \be
FB:\;\; \xi=\fr{z_+-z_-}{\beta (z_++z_-)},\qquad T: v=
\fr{{p}_{+\bot}^2-{p}_{-\bot}^2-\xi { k}_\bot^2}{\beta M|{
k}_\bot|}.\label{asymvar}
 \ee
They can be written via c.m.s. angular variables as
$\xi=\cos\theta$, $v=\sin\theta\cos\phi$.

We describe
 \be
 \Delta\sigma_T = \int d\sigma_{v >0} - \int d\sigma_{v <0}\,,\ \
 \Delta\sigma_{FB} = \int d\sigma_{\xi >0} - \int d\sigma_{\xi <0} \label{defDelta}
 \ee
and $\sigma_{bkgd}=\int d\sigma$ with integration over
(identical) suitable region of final phase space. For the
integrated luminosity ${\cal L}$, the statistical
significance of the result is
 \be
 SS_{T,FB}={\cal L} \Delta\sigma_{T,FB}/\sqrt{{\cal
 L}\,\sigma_{bkgd}}.\label{SSdef}
\ee

$\blacksquare$ {\Large\bf Amplitudes}

Let  $A_-$ and   $A_+$ be amplitudes of production of
C--odd and  C--even dipions with helicities
$\lambda_{\pi\pi}^-$,  $\lambda_{\pi\pi}^+$. Then
 \be
d\sigma\propto  |A_-|^2+|A_+|^2+ 2Re(A_-^*A_+).
 \ee
The interference term $ 2Re(A_-^*A_+)$ describes the charge
asymmetric contribution. At   odd or even
$\lambda_{\pi\pi}^+ -\lambda_{\pi\pi}^-$ we have {\bf T} or
{\bf FB} asymmetry, respectively.

The amplitude $A_-\equiv A_-^\Pom$  is described by Pomeron
exchange. The amplitude $A_+$ is given by the sum of the
photon, odderon and  $\rho/\omega$-Regge exchanges,
$A_+=A_+^\gamma+A_+^{\od}+A_+^{\rho,\omega}$.

We present Pomeron and odderon amplitudes in the form
 \be
{\cal A}_\pm= \sum\limits_{Jn}A_{\pm
Jn}(s,t,M^2)D_J(M^2){\cal
 E}_J^{\lambda_\gamma,\lambda_{\pi\pi}}\;.\label{Pomampl}
 \ee
Here $A_{-Jn}$  and $A_{+Jn}$ are the proper Pomeron and
odderon amplitudes for dipion production,
 $D_J(M)$ describes dipion $\to \pi^+\pi^-$ decay
(e.g. normalized resonance propagator). For Pomeron we use
fit from  \cite{DM2} (with running $\rho$ width and
$\rho'/\rho''$ states), for odderon -- $f_2(1280)$
propagator.  Factor ${\cal
E}^{\lambda_\gamma,\lambda_{\pi\pi}}_J$ describes the
angular distribution of pions; in their c.m.s. frame,
${\cal E}^{\lambda_\gamma,\lambda_{\pi\pi}}_J =
Y^{J,\lambda_{\pi\pi}}(\theta,\phi)e^{-i\lambda_\gamma\psi}$,
where $\psi$ is azimuthal angle of photon polarization
vector ($\psi$ disappears after azimuthal averaging over
momenta of scattered electrons or nuclei).
 The amplitude  $A_+^\gamma$ is the same as in
the $e^+e^-\to e^+e^-\pi^+\pi^-$ and is well known.\\

$\blacksquare$ {\large\bf\boldmath $ep$ and $eA$ collisions
at HERA and e-RHIC }

For the $ep$ collisions at HERA, we take ${\cal
L}_{ep}=100$ pb$^{-1}$, the same values of $SS_T$ are
obtained at e--RHIC with ${\cal L}_{eA}=40$ pb$^{-1}$.

\cl{\large\bf\boldmath $20<k_\bot<100$~MeV, Pomeron phase
\cite{Gin2001}}

Here C--even dipion is produced mainly via the Primakoff
effect -- photon exchange with target ($\gamma\gamma$
collision) ($A_+^{\od},$ $A_+^{\rho,\omega}$ negligible).
Only transverse asymmetry appears due to the $s$--channel
helicity conservation  (SCHC) for Pomeron.
 \begin{figure}
\begin{center}
\epsfig{file=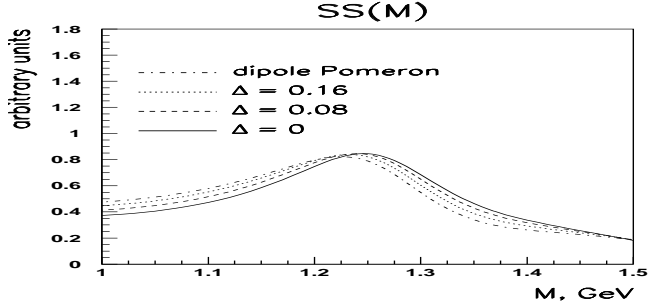,
width=0.8\textwidth,height=4.2cm}
 \caption{The local statistical significance of the charge
asymmetry (arbitrary units).}
\end{center}\vspace{-8mm}
 \end{figure}

Integration over intervals $0.2 < y < 0.8$, $1.1<M<1.4$ GeV
results in
 \be
 \Delta\sigma_{T}\approx 0.13\mbox{ nb} \,,\quad \sigma_{bkgd}^{ep} \approx 1.5\mbox{ nb}
\Rightarrow\;\;{ SS_T \approx 34}.
 \ee

\cl{\large\bf\boldmath Discovery of the odderon, $k_\bot\gtrsim
200$~MeV \cite{Gin2004}}

At  these $k_\bot$ C--even dipion can be produced only via
odderon exchange (photon contribution disappears at these
$k_\bot$; $\rho,\omega$ Regge exchange contributions are
negligibly small $\sigma_{\rho\omega}<0.15$~nb for HERA,
$\sigma_{\rho\omega}<3$~nb for e-RHIC).

We estimated effect, assuming that the $t$-dependence of
odderon amplitude is roughly the same as for Pomeron (but
$M$-dependence is given by $f_2$ contribution). The
shape of local statistical significance in this region is
roughly similar to that for Pomeron--photon interference.
We denote total cross section of $\gamma p\to f_2p$ by
$\sigma_\od$. Different opportunities for helicity of
produced C--even dipion result in different estimates for
asymmetry.

$\nabla$  If SCHC holds also for the odderon amplitude,
the main charge asymmetry is forward--backward. The
integration over $k_\bot>0$ and over $y=0.2-0.8$ results in
 \bear{c}
 \sigma^{ep}_{bkgd}=22\, \mbox{nb},\;\;
\Delta\sigma_{\Pom-\od,FB}=
0.83 \mbox{nb}\,\sqrt{\sigma_\od/\mbox{nb}}\\
\Rightarrow\;{SS_{FB}=56\sqrt{\sigma_\od/\mbox{nb}}>56}
\,.\end{array}\label{chodfb}\ee $\nabla$  If SCHC is
violated strongly for odderon, the main charge asymmetry is
the transverse one, in this case one can integrate over the
region $k_\bot>300$~MeV, to avoid photon exchange
contribution. In this case
 \bear{c}
\sigma^{ep}_{bkgd,T}\approx 9\, \mbox{nb}\,;\quad
\Delta\sigma_{\Pom-\od,T}\approx
0.34\sqrt{\sigma_\od/\mbox{nb}}\, \mbox{nb}\\[2mm]\Rightarrow
{SS_T=35\sqrt{\sigma_\od/\mbox{nb}}>35}.\end{array}
 \label{chodt}\ee
In these equations numbers 35 and 56 correspond
$\sigma_\od=1$~nb.

{\it Therefore, both the measurement of Pomeron phase and
discovery of odderon with high sensitivity are possible at
both colliders.} \\

$\blacksquare$ {\large\bf $A_1A_2\to\pi^+\pi^-A_1A_2$
 at $k_\bot^{\pi\pi}<60$~MeV,
$|k_z^{\pi\pi}|<3$~GeV}

This case corresponds to modern RHIC experiments with $Au$
nuclei. In this kinematical region the two-photon
production of C--even $f_2$ with cross section $\propto
Z^4$ dominates. C--odd dipion is produced in the collision
of almost mass shell photon, radiated e.g. by $A_1$, with
nuclei $A_2$  and {\it vice versa}. The interference of
these amplitudes results in charge asymmetry which changes
sign at transition from $k_z^{\pi\pi}>0$ to
$k_z^{\pi\pi}<0$. The upper limit for $k_\bot$ is given by
nuclear form-factor. Calculations similar to that presented
above show that the statistical significance $SS_T\approx
30$ can be obtained at luminosity integral about 10
nb$^{-1}$. {\it Therefore, the measurement of Pomeron phase
is possible here.} Unfortunately, no definite predictions
can be presented for larger $k_\bot$, and discovery of
odderon
is doubtful.\\

This work was supported by RFBR grant 05-02-16211,
NSh-2339.2003.2.

\end{document}